\documentclass[aps,prb,twocolumn, showpacs]{revtex4}

\usepackage{epsfig}
\usepackage{graphicx}

\def\be{\begin{equation}}
\def\ee{\end{equation}}
\def\bea{\begin{eqnarray}}
\def\eea{\end{eqnarray}}

\begin{document}

\title{Theory of electric dipole spin resonance in quantum dots:\\
Mean field theory with Gaussian fluctuations and beyond}
\author{Emmanuel I. Rashba}
\affiliation{Department of Physics and Center for Nanoscale Systems, Harvard University, Cambridge, Massachusetts 02138, USA\\ 
and Department of Physics, Loughborough University, Leicestershire LE11 3TU, UK}  
\date{\today}

\begin{abstract}
Very recently, the electric dipole spin resonance (EDSR) of single electrons in quantum dots was discovered by three independent experimental groups. Remarkably, these observations revealed three different mechanisms of EDSR: coupling of electron spin to its momentum (spin-orbit), to the operator of its position (inhomogeneous Zeeman coupling), and to the hyperfine Overhauser field of nuclear spins. In this paper, I present a unified microscopic theory of these resonances in quantum dots. A mean field theory, derived for all three mechanisms and based on retaining only two-spin correlators, justifies applying macroscopic description of nuclear polarization to the EDSR theory. In the framework of the mean field theory, a fundamental difference in the time dependence of EDSR inherent of these mechanisms is revealed; it changes from the Rabi-type oscillations to a nearly monotonic growth. The theory provides a regular procedure to account for the higher nuclear-spin correlators that become of importance for a wider time span and can change the asymptotic behavior of EDSR. It also allows revealing the effect of electron spin dynamics on the effective coupling between nuclear spins. 
\end{abstract}

\maketitle

\narrowtext

\section{Introduction}
\label{sec:intro}

One of the principal avenues of semiconductor spintronics is based on the electrical manipulation of electron spins in single and double quantum dots that are envisioned as prospective blocks for quantum computation.\cite{LDV,KouMar98} There are two major aspects of this problem. First is based in electrical control of spin populations by means of the Coulomb and Pauli blockade when electric current passing through the dot populates or depopulates specific spin states.\cite{HansonRMP,Jake} Second aspect is based on direct electrical operation of electron spin by means of the electric dipole spin resonance (EDSR).\cite{R60,RS91} Recently, it was achieved in quantum dots by three different mechanisms.\cite{Nowack07,Laird07,Tarucha08} 

Nowack {\it et al.}\cite{Nowack07} observed Rabi oscillations driven by the electric field generated by {\it ac} gate voltage and coupled to electron spin {\it via} spin-orbit (SO) interaction. Laird {\it et al.}\cite{Laird07} reported hyperfine-mediated gate-driven EDSR remarkable for a nearly monotonic increase of spin polarization rather than Rabi oscillations of it. The underlying mechanism is spacial inhomogeneity of the Overhauser field acting on electron spin. It is physically allied to the EDSR mediated by a spatially dependent Zeeman Hamiltonian, due to the spatial dependence of either the external\cite{PR65,Rashba05JS,Tokura06} or exchange\cite{PR65,KRS} magnetic field or of the Land\'{e} $g$-factor.\cite{KMDGLA03} The SO mechanism may dominate in strong external magnetic fields $B$ but is suppressed in weak fields because the Kramers' theorem requires (in confined geometries) the Rabi frequency to vanish linearly in $B$ as $B\rightarrow0$,\cite{RS64a,Levitov03,Golovach06,ER06} whilst the hyperfine mechanism survives in the $B\rightarrow0$ limit due to the broken time-inversion symmetry and therefore dominates in weak magnetic fields.\cite{SpinRel} More recently, Pioro-Ladri\`{e}re {\it et al.}\cite{Tarucha08} achieved EDSR in a double dot by employing a spatially inhomogeneous (slanting) stray field of a micromagnet and proved high efficiency of this approach. The traditional electron spin resonance (ESR) driven by an {\it ac} magnetic field has also been achieved in quantum dots,\cite{Koppens06} however EDSR promises higher efficiency and provides easier access to individual spins at nanoscale. Golovach {\it et al.} estimated relative intensities of ESR and SO-mediated EDSR in quantum dots and concluded that electron spin can be operated at a timescale of 10 ns.\cite{Golovach06}

The research described in this paper was inspired by the observation of single-electron EDSR in GaAs quantum dots.\cite{Nowack07,Laird07,Tarucha08} Electron spin dynamics in such dots is a challenging problem because of the hyperfine coupling of the electron spin to nuclear spin bath. As applied to spin relaxation in quantum dots, a  mean-field theory approach advanced by Merkulov {\it et al.}\cite{Merkulov02}, Khaetskii {\it et al.},\cite{Khaetskii02} and  Erlingsson and Nazarov\cite{Nazarov02} and based on a large number of nuclear spins in the dots, typically $N\approx10^5-10^6$, proved rather successful. However, some important aspects of electron spin relaxation and dephasing cannot be described in the framework of the mean-field approach, and dynamics of a single electron spin in a nuclear spin environment developed recently into a rather extensive field, e.g., see papers [\onlinecite{DengHu,Witzel06,LuSham06,Coish,Balens,Dobrov08}] and references therein.

Applying the mean field approach to EDSR is highly attractive because it simplifies the problem tremendously. This approach is based on (i) statistical arguments (large number of nuclear spins) and (ii) slow nuclear-spin dynamics. However, large electron spin-flip frequency that keeps nuclear dynamics slow nearly vanishes in the rotating frame\cite{Rabi54,Abraham} under the EDSR conditions. Therefore, the criteria of the applicability of the mean field approach are far from obvious. In particular, the possibility of accelerating spin relaxation in the nuclear bath in the EDSR regime must be examined. We come back to this problem in Sec.~\ref{sec:concl} where it is discussed from the standpoint of the results derived in the paper.

In what follows, I consider different driving forces (SO, Zeeman, and hyperfine mediated), and the hyperfine spin relaxation mechanism that usually dominates in quantum dots, and derive mean field theory equations. It turns out that in all three cases the mean field theory reduces to averaging over the longitudinal and transverse fluctuations of nuclear magnetization, in agreement with intuitive arguments. The results provide a justification for previous theoretical work performed in the framework of this approach. Mathematically, the derivation of mean field theory is based on relating separate terms of the power series for the electron spin flip probability $W(t)$ to Eulerian $\Gamma$-functions of integer or half-integer arguments and employing their integral representations. The theory allows to derive, in the framework of a unified technique, time dependence $W(t)$ of EDSR that is oscillatory for the SO and Zeeman mechanisms, in good agreement with the theory of Koppens {\it et al.},\cite{Koppens07} and shows a nearly monotonic increase for the hyperfine mechanism in accordance with the conclusions by Laird {\it et al.}\cite{Laird07} In the latter case, $W(t)$ approaches its saturation value $W_\infty$ very fast, like a Gaussian exponent, according to the mean field theory, while corrections to this theory show a slow power-law decay.

In this paper, the theory is developed as applied to a single quantum dot; second dot can only serve for probing spin dynamics. The model includes neither electric current across the dot nor the coupling to thermal bath that can result in electrical pumping of nuclear spin polarization and the back action of this magnetization on the electron spin.\cite{Rudner07,Danon08} In electrically pumped double dots this back action is known to result in strong Overhauser fields and instabilities that were observed in the Pauli blockade regime.\cite{Laird07,Koppens06,Tarucha04,Baugh07} For double dots, there is recent progress in the controllable electrical generating nuclear spin polarization\cite{Petta08,Optics} and suppression of electron spin dephasing caused by random fluctuations.\cite{Reilly08} With a dynamically pumped nuclear spin polarization essentially exceeding the polarization from Poissonian fluctuations employed in Ref.~\onlinecite{Laird07} (about 50 G), one can expect dramatic enhancement of the hyperfine-mediated EDSR. I expect that the techniques developed in this paper can be extended to those more general regimes.

The paper is organized as follows. In Sec.~\ref{sec:Ham}, the EDSR Hamiltonians are derived for three mechanisms of the coupling of electron spin to the driving electric field, and afterwards transformed to a form convenient for calculating EDSR probability $W(t)$. A general mean field expression for $W(t)$ is derived in Sec.~\ref{sec:SpFlip}. In Sec.~\ref{sec:SOMM}, the equation for $W(t)$ is simplified as applied to the SO and Zeeman (magnetic) mechanisms of EDSR, and the asymptotic behavior of $W(t)$ is found. In Sec.~\ref{sec:HF}, the same program is performed for a more challenging problem of hyperfine-mediated EDSR. Sec.~\ref{sec:concl} recapitulates the challenges and limitations of the mean field theory as applied to EDSR, and summarizes the basic approaches and results. Nuclear spin relaxation rate in a transient regime is calculated in Appendix A, where absence of its resonance enhancement is shown. Appendix B includes estimates of the corrections to mean field theory.

\section{The Hamiltonian and its transformation}
\label{sec:Ham}

The Hamiltonian that is used in what follows is
\be
H=H_0+H_{el}(t)+H_Z+H_{SO}({\bf k},{\bf r})+H_{hf}({\bf r})\,.
\label{Ham}
\ee
Here $H_0$ is the zero-order Hamiltonian 
\be
H_0={{\hbar^2{\bf k}^2}\over{2m}}+{{m\omega_0^2}\over{2}}{\bf r}^2+V(y)\,,
\label{zero}
\ee
describing an electron moving in $(x,z)$ plane in a parabolic quantum dot, ${\bf r}=(x,z)$, with the in-plane confinement frequency $\omega_0$, being rigidly confined in $y$ direction by the potential $V(y)$. Second term 
\be
H_{el}(t)=e{\tilde{\bf E}}(t)\cdot{\bf r}\,,\,\,{\tilde{\bf E}}(t)=2{\tilde{\bf E}}\cos{\omega t}\,,
\label{el}
\ee
describes the potential energy of the electron, with a charge $(-e)$, in an in-plane driving electric field ${\tilde{\bf E}}(t)\perp{\hat{\bf y}}$. Third term
\be
H_Z=-|g|\mu_B({\bf B}\cdot{\bf s})\,,\,\,{\bf s}=\mbox{\boldmath$\sigma$}/2\,,
\label{Zeeman}
\ee
is the Zeeman energy for an electron with a negative $g$-factor, $g<0$, like in GaAs and InAs, and $\bf s=\mbox{\boldmath$\sigma$}/2$ and $\mbox{\boldmath$\sigma$}$ are the electron spin operator and vector of Pauli matrices, respectively, $\mu_B=e\hbar/2m_0c$ being the Bohr magneton. With a homogeneous field $\bf B$ in the confinement plane, ${\bf B}\perp{\hat{\bf y}}$, as in Refs.~\onlinecite{Nowack07} and \onlinecite{Laird07}, the diamagnetic contribution to the first term in $H_0$ can be disregarded, hence, $\bf k$ will be identified as a canonical momentum.

Generalized SO Hamiltonian $H_{SO}({\bf k},{\bf r})$ consists of two terms, $H_{SO}({\bf k},{\bf r})=H_{SO}({\bf k})+H_{SO}({\bf r})$. Here $H_{SO}({\bf k})$ is the usual momentum-dependent SO interaction, and only the linear in $\bf k$ terms will be considered in what follows. In the geometries of Refs.~\onlinecite{Nowack07} and \onlinecite{Laird07}, the field  ${\tilde{\bf E}}(t)$ was applied along the face-diagonal direction, in our notations along [1,0,1]. In the related coordinate frame, the Rashba and Dresselhaus contributions to $H_{SO}$ can be written as $H_R=\alpha_R(\mbox{\boldmath$\sigma$}\times{\bf k})_y$ and $H_D=\alpha_D(\sigma_xk_z+\sigma_zk_x)$. With $z$ axis chosen along ${\tilde{\bf E}}(t)$, ${\tilde{\bf E}}={\tilde E}{\hat{\bf z}}$, only the $k_z$ component of the momentum $\bf k$ matters in $H_{SO}({\bf k})$. Hence, $H_{SO}({\bf k})$ reduces to a single term 
\be
H_{SO}({\bf k})=\alpha\sigma_xk_z\,,
\label{SO}
\ee
with $\alpha=\alpha_D-\alpha_R$.\cite{DrRa} Also, the homogeneous field ${\bf B}$ will be chosen as ${\bf B}=B{\hat{\bf z}}$ because this is the simplest geometry in which $H_{SO}({\bf k})$ of Eq.~(\ref{SO}) mediates spin flips. 

The term $H_{SO}({\bf r})$ describes SO interaction originating from the inhomogeneity of magnetic field, the mechanism employed in Ref.~\onlinecite{Tarucha08}. With the polarization of the driving field ${\tilde{\bf E}}={\tilde E}{\hat{\bf z}}$, only the $z$-dependence of $H_{SO}({\bf r})$ matters, and to simplify calculations, $H_{SO}({\bf r})$ will be chosen linear in $z$. Keeping only the $y$ component of the stray field,\cite{Tarucha08} one arrives at
\be
H_{SO}({\bf r})=\beta\sigma_yz\,,
\label{SL}
\ee
where $\beta$ is a slanting coefficient. 

The standard expression for the hyperfine Hamiltonian is
\be
H_{hf}({\bf r})=A\sum_j\delta({\bf r}-{\bf r}_j)({\bf I}_j\cdot{\bf s})\,,\,\,A={{16\pi}\over{3I}}\eta\mu_B\mu_n\,,
\label{hf}
\ee
where summation is performed over all lattice sites $j$, ${\bf I}_j$ are operators of nuclear momenta, $I=3/2$ for GaAs, $\mu_n$ are magnetic moments of nuclei (difference in $\mu_n$ values for different nuclei is disregarded), and $\eta$ is the enhancement factor. For GaAs, a rough estimate $An_0\approx10^{-4}$ eV can be used, with $n_0=4.5\times10^{-22}$ cm$^{-3}$ for the concentration of nuclei.\cite{Paget77}

\subsection{Transformation into the moving-dot frame}
\label{sec:moving}

In what follows, $H_{SO}({\bf k},{\bf r})$ and $H_{hf}({\bf r})$ will be considered as small compared with $\hbar\omega_0$. However,  before applying perturbation theory, it is convenient to eliminate the zero mode inherent in the problem of a parabolic dot in a homogeneous field ${\tilde{\bf E}}(t)$. This mode manifests itself in keeping the shape of the electron cloud unchanged when it is displaced by an external homogeneous and time-independent electric field.\cite{DotVsImp} This can be conveniently achieved by performing a time-dependent canonical transformation of the Schroedinger equation $i\hbar\partial_t\Psi=H\Psi$ as
\be
\Psi({\bf r},y,t)\rightarrow{\bf e}^{-i{\bf k}\cdot{\bf R}(t)}{\Psi}({\bf r},y,t),{\bf R}(t)=-e{\tilde{\bf E}}(t)/m\omega_0^2\,;
\label{canon}
\ee
it describes changing to a coordinate frame moving with the dot. This choice of ${\bf R}(t)$ allows eliminating the driving term $H_{el}(t)$ and the zero mode. The transformation of Eq.~(\ref{canon}) also produces a term $-e^2{\tilde E}^2(t)/2m\omega_0^2$ that has no physical significance and can be eliminated by an additional canonical transformation. However, two different results of the transformation to the moving-dot frame have important consequences. 

First, instead of the term $H_{el}(t)$ a new term ${\tilde H}_{el}(t)$ appears in the Hamiltonian
\be
{\tilde H}_{el}(t)=-2{{e\hbar\omega}\over{m\omega_0^2}}({\bf k}\cdot{\tilde{\bf E}})\sin{\omega t}\,.
\label{NewTerm}
\ee
This term, in conjunction with $H_{SO}({\bf k})$, drives the SO mediated EDSR. A factor $\omega/\omega_0\ll1$ in Eq.~(\ref{NewTerm}) emphasizes a critical role of nonadiabaticity for this type of SO coupling and relates the EDSR intensity to it. If the spin resonance frequency $\omega_s=|g|\mu_BB/\hbar$ is small compared with $\omega_0$, as is typical of GaAs, then $\omega/\omega_0\approx\omega_s/\omega_0\ll1$. This is the special form in which the EDSR suppression due to the Kramers theorem\cite{RS64a,Levitov03,Golovach06,ER06} manifests itself as applied to parabolic quantum dots.

Second, the canonical transformation of Eq.~(\ref{canon}) changes the operator $\bf r$ as
\be
{\bf r}\rightarrow e^{i{\bf k}\cdot{{\bf R}(t)}}{\bf r}e^{-i{\bf k}\cdot{{\bf R}(t)}}={\bf r}+{\bf R}(t)\,.
\label{coord}
\ee
This time-dependent shift of $\bf r$ by ${\bf R}(t)$ transforms the $\bf r$-dependent operators $H_{SO}({\bf r})$ and $H_{hf}({\bf r})$ as $H_{SO}({\bf r})\rightarrow H_{SO}({\bf r}+{\bf R}(t))$ and $H_{hf}({\bf r})\rightarrow H_{hf}({\bf r}+{\bf R}(t))$. As applied to $H_{SO}({\bf r})$ of Eq.~(\ref{SL}), the time independent term $\beta\sigma_yz$ produces EDSR only in conjunction with ${\tilde H}_{el}(t)$, hence, this contribution is suppressed by the nonadiabaticity factor $\omega/\omega_0$ and will be omitted. The dominant term in the transformed $H_{SO}({\bf r})$ comes from the $z$-component of ${\bf R}(t)$ and equals
\be
H_{m}(t)=\beta\sigma_yZ(t)\,.
\label{magn}
\ee
It is not subject to the Kramers suppression because of the breaking time inversion symmetry.

\subsection{Projecting onto the ground state}
\label{sec:project}

[sec:project] At this moment, it is convenient to project the moving frame Hamiltonian onto the oscillator-type ground state of the zero-order Hamiltonian $H_0$. Two terms in the total Hamiltonian $H$, $H_{SO}({\bf k})$ and ${\tilde H}_{el}(t)$, are nondiagonal in oscillator quantum numbers. Projecting them onto the oscillator ground state requires performing a standard (Luttinger-Kohn\cite{LK55} or Schrieffer-Wolff\cite{SW66} type) canonical transformation\cite{LK55}
\be
(H_0+H_1)\rightarrow e^T(H_0+H_1) e^{-T}
\label{LK}
\ee
with $H_1=H_{SO}({\bf k})+{\tilde H}_{el}(t)$. Choosing $T$ from the condition of canceling the linear in $H_1$ term, $H_1+[T,H_0]=0$, the Hamiltonian reduces to $(H_0+H_1)\rightarrow H_0-[T,[T,H_0]]/2$ in the quadratic in $T$ approximation. Solving the equation for $T$ results in $\langle 0|T|1\rangle=-\langle0|H_1|1\rangle/\hbar\omega_0$, where $|0\rangle$ and $|1\rangle$ are standing for the oscillator ground and first excited state, respectively. Then, using Eqs.~(\ref{SO}) and (\ref{NewTerm}) and keeping in the second order correction only the term oscillating at the frequency $\omega$, one arrives at a Hamiltonian $H_0+H_{SO}(t)$ with
\be
H_{SO}(t)=2{{\alpha e{\tilde E}}\over{\hbar\omega_0}}{{\omega}\over{\omega_0}}\sigma_x\sin{\omega t}=
2{{e{\tilde E}r_0^2}\over{\ell_{SO}}}{{\omega}\over{\omega_0}}\sigma_x\sin{\omega t}\,.
\label{SOfinal}
\ee
This operator describes the joint effect of $H_{SO}({\bf k})$ and ${\tilde H}_{el}(t)$. In the projected Hamiltonian $H_0=\hbar\omega_0/2$; this constant having no effect on spin dynamics will be omitted.  When deriving Eq.~(\ref{SOfinal}), the expression $|\langle0|k_z|1\rangle|^2=m\omega_0/2\hbar$ for the matrix element of the momentum was used. In this equation $r_0=\sqrt{\hbar/m\omega_0}$ is the electron ground-state radius, and  $\ell_{SO}=\hbar^2/m\alpha$ is a characteristic SO length. Eq.~(\ref{SOfinal}) holds when $\epsilon_\alpha, e{\tilde E}r_0(\omega/\omega_0)\ll\hbar\omega_0$, with $\epsilon_\alpha=m\alpha^2/\hbar^2$ for the characteristic SO energy.

Averaging $H_{hf}({\bf r}+{\bf R}(t))$ over the ground state $\Psi_0({\bf r},y)$ of the dot, expanding the average in ${\bf R}(t)$ and keeping two leading terms of the expansion, results in two hyperfine contributions to the Hamiltonian 
\bea
H_{hf}^0&=&A\sum_j\Psi_0^2({\bf r}_j,y_j)({\bf I}_j\cdot{\bf s})\,,\nonumber\\
H_{hf}(t)&=&A\sum_j({\bf R}(t)\cdot\nabla_{{\bf r}_j})\Psi_0^2({\bf r}_j,y_j)({\bf I}_j\cdot{\bf s})\,;
\label{hfFinal}
\eea
here and below $\Psi_0({\bf r},y)$ is chosen real. The criterion of this expansion, $Z(t)\ll r_0$, is equivalent to $e{\tilde E}r_0\ll\hbar\omega_0$.

The term $H_{hf}^0$ is time independent and describes the hyperfine corrections to the electron Zeeman splitting in the field ${\bf B}=B{\hat{\bf z}}$ and the electron-mediated coupling between nuclear spins. It will be discussed in Sec.~\ref{sec:spsp} in more detail.

The term $H_{hf}(t)$ oscillates at the frequency $\omega$ and results in hyperfine-mediated EDSR. Below, only step-up and step-down spin operators responsible for these transitions will be retained in $H_{hf}(t)$, hence, it reduces to the form
\be
H_{hf}^\perp(t)={A\over4}\sum_jZ(t)\cdot\partial_{z_j}[\Psi_0^2({\bf r}_j,y_j)](I_j^+\sigma_-+I_j^-\sigma_+)\,,
\label{hfTrans}
\ee
where $\sigma_\pm=\sigma_x\pm i\sigma_y$, and $I_j^\pm=I_j^x\pm iI_j^y$.

Finally, the Hamiltonian in the quantum-dot reference frame reads
\be
H=[H_Z+H_{hf}^0]+[H_{SO}(t)+H_m(t)+H_{hf}^\perp(t)]\,.
\label{DotFr}
\ee
Three terms in second bracket represent three EDSR mechanisms discussed above. In deriving this Hamiltonian, only linear in ${\bf R}(t)$ terms were retained. Terms of the higher order in ${\bf R}(t)$, as well as higher harmonics in Eq.~(\ref{SOfinal}) and some different terms, require a theory including parametric excitations.\cite{Walls07} 

\subsection{Spin-spin coupling operator}
\label{sec:spsp}

In this paper, electron spin coupling to the nuclear spin bath and the spectral diffusion caused by this coupling are considered as the only source of electron spin decoherence because of the  general scope of the paper focused on the hyperfine coupling and especially on the interplay between its contributions to different terms of the perturbation theory. This decoherence mechanism was recently discussed in the context of free spin induction decay and Hahn echo by Witzel and Das Sarma\cite{Witzel06} and Yao, Liu, and Sham;\cite{LuSham06} see also references in these papers to extensive literature on the subject. Electron spin decoherence originates from the spin dynamics in the nuclear bath caused by both the intrinsic mechanisms (like dipole-dipole interaction) and the extrinsic mechanism mediated by the hyperfine coupling of nuclear spins to the electron spin through the operator $H_{hf}^0$ of Eq.~(\ref{hfFinal}). In what follows, only the latter mechanism will be considered because it facilitates effective nuclear spin-spin coupling at the large spatial scale of $r_0$.

The term $H_{hf}^0=H_{hf}^\parallel+H_{hf}^\perp$ includes two contributions. The longitudinal (secular) part $H_{hf}^\parallel$ is proportional to $s_z$ and describes random fluctuations of the electron Zeeman energy because of the Overhauser field. The transverse part $H_{hf}^\perp$, while depending on $s_\pm$, cannot produce real electron spin flip transitions because of the large electron Zeeman energy $\hbar\omega_s=|g|\mu_BB$ (nuclear Zeeman energy is small and will be omitted). However, in the second order of the perturbation theory in $\omega_s^{-1}$, the operator $H_{hf}^\perp$ results in nuclear spin nonconservation and spectral diffusion of the Overhauser field. Everywhere below, nuclear spins ${\bf I}_j$ will be considered as classical variables, i.e., their commutators will be disregarded. This simplifies calculations and should not affect basic results because ${\bf I}_j$ and ${\bf I}_{j'}$ commute for $j\neq j'$, and correlators involving more than two spin operators at the same lattice site are statistically insignificant in the mean field theory developed below (Appendix B is the only exclusion).

Applying the procedure of Eq.~(\ref{LK}) to the operator $H_Z+H_{hf}^\perp$, one finds $T=\sum_ja_j(I_j^+\sigma_--I_j^-\sigma_+)/\sqrt{8\hbar\omega_s}$, with
\be
a_j={{A}\over{\sqrt{2\hbar\omega_s}}}\Psi_0^2({\bf r}_j,y_j)\,,
\label{aDefin}
\ee
and the transformed operator $H_{hf}^\perp$ becomes an operator of the effective nuclear spin-spin coupling
\be
H_{ss}=-{1\over2}\sum_{n\neq m}a_na_mI_n^+I_m^-\sigma_z\,.
\label{spsp1}
\ee
It interchanges the projections of nuclear spins at different lattice sites while keeping electron spin unchanged. This expression for spin-spin coupling recovers the result by Yao {\it et al.}\cite{LuSham06} Because the diagonal term, $n=m$, is statistically insignificant for large dots, the $n\neq m$ constraint can be omitted. Then 
\be
H_{ss}=-{1\over2}~g_+g_-\sigma_z\,,\,\,g_\pm=\sum_ja_jI_j^\pm\,.
\label{spsp}
\ee
In similar notations
\be
H_{hf}^\parallel=g_z\sigma_z\,,\,\,g_z=\sqrt{\hbar\omega_s/2}~\sum_ja_jI_j^z\,.
\label{parall}
\ee
We notice that $g_z$ and $g_\pm$ have different dimensions.

Substituting (\ref{spsp}) and (\ref{parall}) into (\ref{DotFr}), one arrives at the final form of the Hamiltonian in the dot frame
\bea
H&=&[H_Z+H_{hf}^\parallel+H_{ss}]\nonumber\\
&+&[H_{SO}(t)+H_m(t)+H_{hf}^\perp(t)]
\label{LabFr}
\eea
Three terms in the first bracket describe the electron Zeeman energy, its random shift due to the longitudinal component of the Overhauser field, and coupling between nuclear spins, respectively. Second bracket describes the driving force acting on the electron spin. It consists of the spin-orbit, inhomogeneous magnetic field, and hyperfine contributions.

In conclusion of this section, one comment regarding the spin-spin Hamiltonian $H_{ss}$ of Eq.~(\ref{spsp1}) should be made. It was derived by transforming the static Hamiltonian $H_Z+H_{hf}^\perp$, and the products $a_na_m$ in $H_{ss}$ are proportional to $1/\omega_s$, with $\omega_s$ playing a role of a large parameter (spin gap) in the energy spectrum. However, in the EDSR regime the situation changes drastically because in the rotating frame Hamiltonian the large frequency $\omega_s$ is reduced to $(\omega_s-\omega)$, see Eq.~(\ref{RF3}) below. Therefore, a question arises whether $H_{ss}$ can experience a resonant enhancement at $\omega\rightarrow\omega_s$ because of the ``rotating frame singularity". This problem is considered in Appendix A, where it is shown that $H_{ss}$ changes and becomes time dependent at the scale of the Rabi frequency but does not experience any resonant enhancement. Because solving electron spin dynamics with a time dependent $H_{ss}(t)$ is an extremely challenging (or even impossible) task, in what follows the Hamiltonian $H_{ss}$ of Eq.~(\ref{spsp}) is used as a model Hamiltonian. The magnitude of the coefficients $a_j$ can be subject to renormalization, but their $j$-dependence will be chosen according to Eq.~(\ref{aDefin}).

\subsection{Rotating frame Hamiltonian}
\label{sec:RotFrame}

All terms in second bracket of Eq.~(\ref{LabFr}) depend on time harmonically, $H_{SO}(t)$ as $\sin{\omega t}$ while $H_m(t)$ and $H_{hf}^\perp(t)$ as $\cos{\omega t}$. Applying the standard transformation to the rotating frame\cite{Rabi54,Abraham}
\be
\Psi({\bf r},y,t)=\exp{(i\sigma_z\omega t/2)}\Psi_{RF}({\bf r},y,t)\,,
\label{RF1}
\ee
that transforms Pauli matrices as
\be
\sigma_\pm(t)=\sigma_\pm\exp{(\mp i\omega t)}\,,\,\sigma_z(t)=\sigma_z\,,
\label{RF2}
\ee
results in the rotating frame Hamiltonian
\be
{\hat H}=\bigg({1\over2}\Delta_s+g_z-{1\over2}~g_+g_-\bigg)\sigma_z+{1\over2}(f_-\sigma_++f_+\sigma_-)\,.
\label{RF3}
\ee
When deriving (\ref{RF3}), the rotating wave approximation was applied with all fast oscillating terms omitted. Here  $\Delta_s=\hbar(\omega-\omega_s)$ is detuning, $g_z$ is the Overhauser shift, and $g_+g_-/2$ describes second order coupling to the spin bath.

Using (\ref{magn}), (\ref{SOfinal}), and (\ref{hfTrans}), one finds driving terms $f_\pm$ for different mechanisms of EDSR
\be
f_\pm^{SO}=\pm i(e{\tilde E}r_0^2/\ell_{SO})(\omega/\omega_0)\,,
\label{RFSO}
\ee
\be
f_\pm^m=\mp i\beta(e{\tilde E}/m\omega_0^2)\,,
\label{RFm}
\ee
\be
f_\pm^{hf}=\sum_jb_jI_j^\pm\,,\,\,b_j=-{{Ae{\tilde E}}\over{2m\omega_0^2}}\partial_{z_j}[\Psi_0^2({\bf r}_j,y_j)]\,.
\label{RFhf}
\ee
It is seen from Eqs.~(\ref{RFSO}) - (\ref{RFhf}) that for two first mechanisms the expressions for $f_\pm$ are very similar, hence, a unified theory of EDSR will be developed for them, see Sec.~\ref{sec:SOMM}. As distinct from them, presence of spin angular momenta $I_j^\pm$ in the coefficients $f^{hf}_\pm$ for the hyperfine mediated EDSR changes the situation drastically. This mechanism requires a special consideration, and a theory for this type of EDSR is described in Sec.~\ref{sec:HF}.

Eq.~(\ref{RF3}) for the Hamiltonian ${\hat H}$, equations (\ref{spsp}) - (\ref{parall}) for $(g_z,g_\pm)$, and (\ref{RFSO}) - (\ref{RFhf}) for $f_\pm$ form the basic system of equations. In what follows, they will be solved and spin flip probabilities will be averaged over nuclear spin configurations for calculating time dependence of EDSR.

\section{Spin flip probability}
\label{sec:SpFlip}

With the external magnetic field $B$ strong enough, the initial electron spin state can be chosen as $|\uparrow\rangle$, and electron wave function in the rotating frame evolves as $\Psi_{RF}(t)=\exp{(-i{\hat H}t/\hbar)}|\Psi_0\uparrow\rangle$. Then, applying (\ref{RF1}) results in the spin flip matrix element
\[\langle\downarrow\Psi_0|\exp{(i\sigma_z\omega t/2)}\exp{(-i{\hat H}t/\hbar)}|\Psi_0\uparrow\rangle\]
and spin flip probability
\be
W(t)=|\langle\downarrow\Psi_0|\exp{(-i{\hat H}t/\hbar)}|\Psi_0\uparrow\rangle|^2\,,
\label{SFP1}
\ee
because the factor $\exp{(i\sigma_z\omega t/2)}$ cancels out. The square of the Hamiltonian $\hat H$ of Eq.~(\ref{RF3})
\be
{\hat H}^2\equiv {\cal H}^2=\bigg({{\Delta_s}\over{2}}+g_z-{1\over2}~g_+g_-\bigg)^2+f_+f_-
\label{SFP2}
\ee
does not depend on Pauli matrices. Hence, the exponential factor in (\ref{SFP1}) can be simplified as
\be
\exp(-i{\hat H}t/\hbar)=\cos({\cal H}t/\hbar)-i({\hat H}/{\cal H})\sin({\cal H}t/\hbar)\,.
\label{SFP3}
\ee
Finally, the transition probability equals
\be
W(t)=\bigg|\bigg\langle\Psi_0\bigg|{{f_+f_-}\over{{\cal H}^2}}\sin^2{({\cal H}t/\hbar)}\bigg|\Psi_0\bigg\rangle\bigg|^2\,.
\label{SFP4}
\ee
This is the celebrated Rabi formula with $f_+f_-$ for the driving term and $(\Delta_s+2g_z)$ for detuning.\cite{Rabi37} The special feature of Eq.~(\ref{SFP4}) is presence of the dephasing term $-g_+g_-/2$ originating from the transverse part of the random nuclear magnetization.

Spin oscillations described by $W(t)$ are controlled by the competition between the quantities $\Delta_s$, $g_z$, $g_+g_-$, and $f_+f_-$, all of them being small compared with the Zeeman energy $H_Z$ that does not appear in (\ref{SFP4}) explicitly. Corrections to the initial state $\Psi_0|\uparrow\rangle$ due to the nuclear spin fluctuations should merely renormalize the basic parameters, in inverse powers of $H_Z$, similarly to the Bloch-Siegert corrections to the rotating wave approximation,\cite{BlochSieg} without changing the main pattern of oscillations. Hence, they will be disregarded in what follows.

\subsection{Averaging over nuclear angular momenta}
\label{sec:PairCor}

The probability $W(t)$ depends on the nuclear spin polarization through $g_z$ and $g_\pm$ of Eqs.~(\ref{spsp}) - (\ref{parall}) and $f^{hf}_\pm$ of (\ref{RFhf}). Because experimental data are typically taken by averaging over dozens of thousand pulses covering time spans exceeding the nuclear spin diffusion time,\cite{Nowack07,Laird07,Tarucha08} average values of $W(t)$ of Eq.~(\ref{SFP4}) over all nuclear spin configurations are of the principal interest. Calculating these average values is highly facilitated by the fact that the series for the ratio
\be
{{\sin({\cal H}t/\hbar)}\over{\cal H}}={t\over\hbar}\sum_{k=0}^\infty(-)^k{{({\cal H}t/\hbar)^{2k}}\over{(2k+1)!}}
\label{AV1}
\ee
includes only even powers of $\cal H$. Therefore, each term of the power series for the averaged probability $W(t)$
\bea
&&W(t)=\sum_{k,k^\prime=0}^\infty{{(-)^{k+k^\prime}(t/\hbar)^{2(k+k^\prime+1)}}\over{(2k+1)!(2k^\prime+1)!}}\nonumber\\
&\times&\bigg\langle f_+f_-\bigg[\bigg({{\Delta_s}\over{2}}+g_z-{1\over2}g_+g_-\bigg)^2+f_+f_-\bigg]^{k+k^\prime}\bigg\rangle_{0,{\rm nuc}}
\label{AV2}
\eea
is a polynomial in the nuclear angular momenta ${\bf I}_j$. The subscript $\{0,{\rm nuc}\}$ indicates that both quantum averaging over $\Psi_0$ and statistical averaging over nuclear momenta should be performed. Due to the weak interaction between nuclear momenta, only single-site correlators can be retained. Next simplification originates from the fact that for large quanum dots, containing about $10^5-10^6$ nuclear spins, pair correlators statistically dominate; the contribution from higher correlators is estimated in Appendix B. Finally, of all pair correlators only $\langle I_j^zI_j^z\rangle=I(I+1)/3$ and $\langle I_j^+I_j^-\rangle=2I(I+1)/3$ do not vanish. This allows to separate the averaging over $g_z$ from the averaging over $g_\pm$ and $f_\pm$.

\subsection{Averaging over longitudinal magnetization}
\label{sec:Long}

From Eqs.~(\ref{aDefin}) and (\ref{parall}) follows an expression for $g_z^2$ averaged over nuclear spin configurations
\be
\langle g_z^2\rangle={1\over{12}}A^2n_0I(I+1)\int\int d{\bf r}~dy~\Psi_0^4({\bf r},y)\,.
\label{Lon1}
\ee
With oscillator ground-state functions $\psi_0(x)=\exp{(-x^2/2r_0^2)}/\sqrt{\pi^{1/2}r_0}$ and $\psi_0(z)$ in $x$ and $z$ directions, and hard wall confinement $\psi(y)=\sqrt{2/d}~\cos{(\pi y/d)}$ in $y$ direction, $\Psi_0({\bf r},y)=\psi_0(x)\psi(y)\psi_0(z)$ and
\be
\langle g_z^2\rangle=\Delta^2/2\,,\,\,\Delta=\sqrt{A^2n_0I(I+1)/8\pi r_0^2d}\,.
\label{Lon2}
\ee
By combinatorial arguments based on the multiplicity of possible pairings, expressing $\langle g_z^{2M}\rangle$ in terms of pair correlators results in
\be
\langle g_z^{2M}\rangle=(2M-1)!!~\langle g_z^2\rangle^M\,,
\label{Lon3}
\ee
with $(2M-1)!!=1$ for $M=0$. Then, from the relation between $(2M-1)!!$ and the Eulerian Gamma function
\be
(2M-1)!!=2^M\Gamma(M+1/2)/\sqrt{\pi}\,,
\label{Lon4}
\ee
and from the integral representation of $\Gamma(M+1/2)$ 
\be
\Gamma\bigg(M+{1\over2}\bigg)=\int_{-\infty}^\infty dt~ t^{2M}~e^{-t^2}
\label{Lon5}
\ee
valid for integer values of $M$, a Gaussian distribution for $g_z$ follows
\bea
\langle g_z^{2M}\rangle&=&\int_{-\infty}^\infty dw~w^{2M}\rho_\Delta(w)\,,\nonumber\\
\rho_\Delta(w)&=&{{1}}\over{\Delta\sqrt{\pi}}}\exp{(-w^2/\Delta^2)\,,
\label{Lon6}
\eea
with a standard deviation $\langle g_z^2\rangle=\Delta^2/2$.

Therefore, averaging (\ref{AV2}) over the longitudinal magnetization $g_z$ can be performed as a Gaussian integration with the distribution function $\rho_\Delta(w)$ of Eq.~(\ref{Lon6}). 

\subsection{Averaging over transverse magnetization}
\label{sec:Trans}

Averaging Eq.~(\ref{AV2}) over $I_j^\pm$ is more involved because these operators appear both in $g_\pm$ and $f_\pm$. Applying 
(\ref{aDefin}) and (\ref{spsp}) results in an expression similar to (\ref{Lon1}), and finally results in
\be
\langle g_+g_-\rangle={{A^2n_0I(I+1)}\over{4\pi\hbar\omega_sr_0^2d}}\,.
\label{Tr1}
\ee
Average values of $(g_+g_-)^M$, in the pair correlator approximation, calculated from combinatorial arguments, are
\be
\langle(g_+g_-)^M\rangle=M!~\langle g_+g_-\rangle^M\equiv M!~\Xi^{2M}\,.
\label{Tr2}
\ee
From the integral representation
\be
M!=\Gamma(M+1)=\int_0^\infty dt~t^Me^{-t}\,,
\label{Tr3}
\ee
by changing the variable $t=u^2/\Xi^2$, one arrives at a Gaussian distribution
\bea
\langle(g_+g_-)^M\rangle&=&\int d^2{\bf u}~u^{2M}\rho_{\Xi}(u)\,,\nonumber\\
\rho_{\Xi}(u)&=&{{1}\over{\pi\Xi^2}}\exp{(-u^2/{\Xi}^2})\,.
\label{Tr4}
\eea
Not surprisingly, the distribution is two-dimensional. This reflects the two-dimensional nature of the transverse spin polarization $I_j^\pm$. Constants $\Delta$ and $\Xi$ are connected by a simple relation
\be
\Delta=\sqrt{\hbar\omega_s/2}~\Xi\,.
\label{Tr5}
\ee
Condition of spin resonance, $\Delta\ll\hbar\omega_s$, is tantamount to
\be
\Xi^2/2\Delta\ll1\,.
\label{Tr5a}
\ee
This condition of weak spin-spin coupling will be used below.

Next step is averaging (\ref{AV2}) over $f_\pm^{hf}$. It follows from (\ref{RFhf}) that
\bea
\langle f_+^{hf}f_-^{hf}\rangle&=&{2\over3}n_0I(I+1)\bigg({{Ae{\tilde E}}\over{2m\omega_0^2}}\bigg)^2\nonumber\\
&\times&\int\int d{\bf r}~dy~\bigg\{\partial_z[\Psi_0^2({\bf r},y)]\bigg\}^2\,,
\label{Tr6}
\eea
and after performing all integrations
\be
\langle f_+^{hf}f_-^{hf}\rangle={{n_0I(I+1)}\over{8\pi d}}\bigg({{Ae{\tilde E}}\over{\hbar\omega_0}}\bigg)^2\equiv\Omega^2\,.
\label{Tr7}
\ee
The same combinatorial arguments that led to (\ref{Tr2}) - (\ref{Tr4}) result in a two-dimensional Gaussian distribution
\bea
\langle(f_+^{hf}f_-^{hf})^M\rangle&=&\int d^2{\bf v}~v^{2M}\rho_\Omega(v)\,,\nonumber\\
\rho_\Omega(v)&=&{{1}}\over{\pi\Omega^2}}\exp{(-v^2/\Omega^2)\,.
\label{Tr8}
\eea

The last sort of pair products appearing in (\ref{AV2}) are cross-products $g_+f_-^{hf}$ and $g_-f_+^{hf}$. According to (\ref{aDefin}), (\ref{spsp}), and (\ref{RFhf})
\be
\langle g_+f_-^{hf}\rangle\propto\int\int d{\bf r}~dy~\partial_z[\Psi_0^4({\bf r},y)]=0\,,
\label{Tr9}
\ee
hence, cross-products vanish after the integration. Vanishing the cross-products allows separating the averaging over the products $g_+g_-$ from averating over $f_+^{hf}f_-^{hf}$. Potentially, cross-products describe the effect of driving force on spin dynamics of the nuclear bath, but in the approximation accepted their average values vanish.

\subsection{Mean field equation}
\label{sec:Mean}

It was shown in Secs.~\ref{sec:Long} and \ref{sec:Trans} that averaging over nuclear spins reduces to three integrations over the distributions $\rho_\Delta(w)$, $\rho_{\Xi}(u)$, and $\rho_\Omega(v)$. For these integrations, the variables in (\ref{AV2}) should be substituted as
\[ g_z\rightarrow w,\, g_+g_-\rightarrow u^2,\, f_+f_-\rightarrow v^2\,.\]
Afterwards, both summations over $k$ and $k^\prime$ in (\ref{AV2}) can be performed, and one recovers the Rabi formula. However, now its arguments are auxiliary variables $\bf u$, $\bf v$, and $w$, over which integration should be performed. Finally
\bea
W(t)&=&\int_{-\infty}^\infty dw~\rho_\Delta(w)\int d^2{\bf v}~\rho_\Omega(v)\int d^2{\bf u}~\rho_{\Xi}(u)\nonumber\\
&\times&v^2{{\sin^2\bigg(\sqrt{\bigg({{\Delta_s}\over{2}}+w-{1\over2}u^2\bigg)^2+v^2}~{{t}\over{\hbar}}\bigg)}
\over{\bigg({{\Delta_s}\over{2}}+w-{1\over2}u^2\bigg)^2+v^2}}\,,
\label{Mean1}
\eea
with $\rho_\Delta(w)$, $\rho_{\Xi}(u)$, and $\rho_\Omega(v)$ defined by (\ref{Lon6}), (\ref{Tr4}), and (\ref{Tr8}), respectively. Eq.~(\ref{Mean1}) generalizes the equation applied in Ref.~\onlinecite{Laird07} to hyperfine-mediated EDSR by including nuclear spin dynamics described by the term $u^2/2$ and the corresponding integration.

The above derivation proves that as applied to EDSR the pair-correlator approximation is tantamount to a mean field theory described by Eq.~(\ref{Mean1}). It was derived for the hyperfine-mediated EDSR. For the spin-orbit or magnetically mediated EDSR the integration over $\bf v$ should be omitted and $v^2$ should be substituted by $f_+^{SO}f_-^{SO}$ or $f_+^mf_-^m$, respectively. Corrections to the mean field theory are estimated in Appendix B.

\section{Spin-orbit and magnetically mediated EDSR}
\label{sec:SOMM}

In this case (\ref{Mean1}) simplifies and includes integration two variables, $\bf u$ and $w$, with $v^2$ substituted as $v^2\rightarrow f^2=f_+f_-$ with $f_+^{SO}f_-^{SO}$ or $f_+^mf_-^m$ of Eqs.~(\ref{RFSO}) and (\ref{RFm}) for the EDSR mediated by the spin-orbit or inhomogeneous magnetic field mechanisms, respectively. The results are also applicable to the usual spin resonance driven by an {\it ac} magnetic field. It is convenient, by using the relation $\sin^2\phi=(1-\cos{2\phi})/2$, to split $W(t)$ onto its asymptotic value $W_\infty$ at $t\rightarrow\infty$ and the time dependent part $W_1(t)$ as
\be
W(t)=W_\infty-W_1(t)\,,
\label{SOM1}
\ee
and introduce dimensionles variables 
\be
\zeta=w/f\,,\,\,\xi=u^2/2f\,,\,\,\tau=ft/\hbar\,,
\label{SOM2}
\ee
and parameters
\be
\delta=\Delta_s/2f\,,\,\,R=f/\Delta\,,\,\,R_1=2f/\Xi^2\,.
\label{SOM3}
\ee
Then
\be
W_\infty={{RR_1}\over{2\sqrt{\pi}}}\int_{-\infty}^\infty d\zeta\int_0^\infty d\xi~
{{\exp{(-R^2\zeta^2-R_1\xi)}}\over{(\delta+\zeta-\xi)^2+1}}\,,
\label{SOM4}
\ee
and
\bea
W_1(\tau)&=&{{RR_1}\over{2\sqrt{\pi}}}\int_{-\infty}^\infty d\zeta~ e^{-R^2\zeta^2}\int_0^\infty d\xi~ e^{-R_1\xi}\nonumber\\
&\times&{{\cos{(2\sqrt{(\delta+\zeta-\xi)^2+1}~\tau)}}\over{(\delta+\zeta-\xi)^2+1}}\,.
\label{SOM5}
\eea

$W_\infty$ can be calculated in the limit of weak driving force, $R\ll1$, due to the criterion of Eq.~(\ref{Tr5a}); this regime is typical of most experiments. Indeed, the integrand includes a narrow Lorentzian $\delta$-function $\delta(\zeta+\delta-\xi)\approx\pi^{-1}/[(\zeta+\delta-\xi)^2+1]$ that allows performing integration over $\zeta$ for $R\ll1$.
Afterwards, integration over $\xi$ can be easily performed because the Gaussian exponent is nearly a constant due to the fact that $2R^2\delta/R_1=(\Delta_s/\Delta)(\Xi^2/2\Delta)\ll1$ for any $\Delta_s/\Delta\alt1$ due to (\ref{Tr5a}). Finally,
\be
W_\infty\approx{{f}\over{2\sqrt{\pi}\Delta}}\exp{\bigg(-{{\Delta_s^2}\over{4\Delta^2}}\bigg)}\,.
\label{SOM6}
\ee
When (\ref{Tr5a}) is satisfied, $W_\infty$ does not depend on $\Xi$. General expression for $W_\infty$ can be found from (\ref{Trans7}) - (\ref{Trans9}) by omitting factors $\cos{(2s\tau)}$ in integrands.

\subsection{Simplifying the expression for $W_1(\tau)$}
\label{sec:siplify}

To simplify integral (\ref{SOM5}), it is convenient to introduce instead of $\xi$ a new variable
\be
s^2=(\delta+\zeta-\xi)^2+1\,,\,\,s\geq0\,,
\label{Trans1}
\ee
which allows expressing $W_1(\tau)$ in terms of a Fourier transformation. Afterwards, integration over $\zeta$ in (\ref{SOM5}) can be cut into parts in such a way that, after simple transformations like changing signs of variables, the integrations are performed over positive values of both variables. Finally, one arrives at
\be
W_1(\tau)=W_1^{(1)}(\tau)+W_1^{(2)}(\tau)+W_1^{(3)}(\tau)
\label{Trans2}
\ee
with
\begin{widetext}
\be
W_1^{(1)}(\tau)={{RR_1}\over{2\sqrt{\pi}}}\int_0^\infty d\zeta\exp{[-R^2(\zeta+\delta)^2+R_1\zeta]}
\int^\infty_{\sqrt{\zeta^2+1}}ds~{{\exp{(-R_1\sqrt{s^2-1})}}\over{s\sqrt{s^2-1}}}~\cos{(2s\tau)}\,,
\label{Trans3}
\ee
\be
W_1^{(2)}(\tau)={{RR_1}\over{2\sqrt{\pi}}}\int_0^\infty d\zeta\exp{[-R^2(\zeta-\delta)^2-R_1\zeta]}
\int_1^{\sqrt{\zeta^2+1}}ds~{{\exp{(R_1\sqrt{s^2-1})}}\over{s\sqrt{s^2-1}}}~\cos{(2s\tau)}\,,
\label{Trans4}
\ee
\be
W_1^{(3)}(\tau)={{RR_1}\over{2\sqrt{\pi}}}\int_0^\infty d\zeta\exp{[-R^2(\zeta-\delta)^2-R_1\zeta]}
\int^\infty_1ds~{{\exp{(-R_1\sqrt{s^2-1})}}\over{s\sqrt{s^2-1}}}~\cos{(2s\tau)}\,.
\label{Trans5}
\ee
\end{widetext}
After introducing two auxiliary functions
\be
G_\pm(\zeta)=\int_\infty^\zeta du~\exp{[-R^2(u\pm\delta)^2\pm R_1u]}\,,
\label{Trans6}
\ee
$G_\pm(\infty)=0$, equations (\ref{Trans3}) - (\ref{Trans4}) can be performed by parts. This transformation reduces repeated integrals to onefold integrals because $\zeta$ enters into internal integrals only through their limits. Finally
\bea
W_1^{(1)}(\tau)&=&{{RR_1}\over{2\sqrt{\pi}}}\int_0^\infty ds~{{G_+(\sqrt{s^2-1})-G_+(0)}\over{s\sqrt{s^2-1}}}\nonumber\\
&\times&\exp{(-R_1\sqrt{s^2-1})}\cos{(2s\tau)}\,,
\label{Trans7}
\eea
\be
W_1^{(2)}(\tau)=-{{RR_1}\over{2\sqrt{\pi}}}
\int_0^\infty ds{{G_-(\sqrt{s^2-1})}\over{s\sqrt{s^2-1}}}e^{R_1\sqrt{s^2-1}}\cos{(2s\tau)},
\label{Trans8}
\ee
\be
W_1^{(3)}(\tau)=-{{RR_1}\over{2\sqrt{\pi}}}\int_0^\infty ds{{G_-(0)}\over{s\sqrt{s^2-1}}}e^{-R_1\sqrt{s^2-1}}\cos{(2s\tau)}.
\label{Trans9}
\ee
Because $G_\pm(\zeta)$ can be expressed in terms of the Erfc-function, integrals (\ref{Trans6}) - (\ref{Trans9}) can be easily calculated for any given set of parameter values. However, a wider insight comes from calculating their asymptotic behavior at large $\tau$.

\subsection{Large $\tau$ behavior of $W_1(\tau)$}
\label{sec:asym1}

Large-$\tau$ behavior of $W_1(\tau)$ is controlled by the singularities of the integrands of the Fourier integrals (\ref{Trans6}) - (\ref{Trans9}). All integrands decay exponentially at $s\rightarrow\infty$, and square-root behavior at $s=1$ is their only singularity. Because the numerator of the integrand of $W_1^{(1)}(\tau)$ vanishes at $s=1$, the singularity is softened. Therefore, the large $\tau$ behavior is controlled by $W_1^{(2)}(\tau)$ and $W_1^{(3)}(\tau)$ due to $\sqrt{s^2-1}$ in the denominators. For calculating the leading terms of the expansions, all other factors in the integrands can be takes at their $s=1$ values, as $G_-(\sqrt{s^2-1})\rightarrow G_-(0)$, $\exp{(\pm R_1\sqrt{s^2-1})}\rightarrow1$, and $s\sqrt{s^2-1}\rightarrow\sqrt{2}\sqrt{s-1}$. The remaining integral reduces to Fresnel integrals and can be easily calculated
\be
\int_1^\infty{{ds}\over{\sqrt{s-1}}}\cos{(2s\tau)}=\sqrt{{\pi}\over{2\tau}}\cos{\bigg(2\tau+{{\pi}\over{4}}\bigg)}
\label{Asym1}
\ee

The coefficient $G_-(0)$ equals
\bea
G_-(0)&=&-{{\sqrt{\pi}}\over{2R}}\exp{\bigg({{R_1^2}\over{4R^2}}-R_1\delta\bigg)}{\rm Erfc}\bigg({{R_1}\over{2R}}-R\delta\bigg)\nonumber\\
&\approx&R_1^{-1}\exp{(-\Delta_s^2/4\Delta^2)}\,;
\label{Asym2}
\eea
second part of (\ref{Asym2}) follows from (\ref{Tr5a}). Remarkably, while exact equation is asymmetric in $\delta$, the approximate expression is symmetric in the detuning $\Delta_s$. Finally, for $t\rightarrow\infty$
\be
W_1(t)\approx{{1}\over{2\Delta}}\sqrt{{{\hbar f}\over{t}}}\exp{\bigg(-{{\Delta_s^2}\over{4\Delta^2}}\bigg)}\cos{\bigg({{2f}\over{\hbar}}t+{{\pi}\over{4}}\bigg)}\,.
\label{Asym3}
\ee

Eq.~(\ref{Asym3}) was derived in a strict $t\rightarrow\infty$ limit. Because of the existence in the problem of a large parameter $R_1\gg1,R,R^{-1}$, one cannot exclude the existence of some intermediate asymptotic valid for $\tau$ up to some upper bound related to this parameter. To check this possibility, one must first take the limit $R_1\rightarrow\infty$ in (\ref{SOM5}). Then 
\bea
W_1(\tau)&=&{{R}\over{\sqrt{\pi}}}\int_1^\infty {{ds}\over{s\sqrt{s^2-1}}}\cosh{(2R^2\delta\sqrt{s^2-1})}\nonumber\\
&\times&\exp{[-R^2(s^2-1+\delta^2)]}\cos{(2s\tau)}\,.
\label{Asym4}
\eea
Applying an approach similar to that leading to (\ref{Asym1}), one arrives at (\ref{Asym3}). Therefore, it holds in the whole region $1\alt\tau<\infty$. This type of universal oscillatory behavior with $1/\sqrt{t}$ decay was first found in the $R_1\rightarrow\infty$ limit, experimentally and theoretically, by Koppens {\it et al.}.\cite{Koppens07}

\section{Hyperfine-mediated EDSR}
\label{sec:HF}

To evaluate the three-fold integral of Eq.~(\ref{Mean1}), it is convenient to split it into two terms similarly to (\ref{SOM1}) and define dimensionless variables 
\be
\zeta=w/\Omega\,,\,\,\eta=v/\Omega\,,\,\,\xi=u^2/2\Omega\,,\,\,\tau=\Omega t/\hbar
\label{HFM1}
\ee
and parameters
\be
R=\Omega/\Delta\,,\,\,R_1=2\Omega/\Xi^2\,,\,\,\delta=\Delta_s/2\Omega\,.
\label{HFM2}
\ee
Then 
\bea
W_\infty&=&{{RR_1}\over{\pi}}\int_{-\infty}^\infty d\zeta~e^{-R^2\zeta^2}\int_0^\infty d\xi~e^{-R_1\xi}\nonumber\\
&\times&\int_0^\infty d\eta~\eta^3~{{\exp{(-\eta^2)}}\over{(\delta+\zeta-\xi)^2+\eta^2}}\,,
\label{HFM3}
\eea
and this integral can be evaluated in the $R\ll1$ limit by performing integration over $\zeta$. Using (\ref{Tr5a}) one arrives at
\be
W_\infty\approx(\pi \Omega/4\Delta)\exp{(-\Delta_s^2/4\Delta^2)}
\label{HFM4}
\ee
that is similar to (\ref{SOM6}). More general expression for $W_\infty$ can be found from (\ref{As3}) by omitting  $\cos{(2\rho\tau)}$ in the integrand.

Evaluating $W_1(\tau)$ is the major challenge, and rewriting it as a sum of three integrals similar to (\ref{Trans2}) (each one including integrations only over positive values of all arguments) is the first step. Transformations are similar to the applied when deriving (\ref{Trans3}) - (\ref{Trans5}), but for the reasons that will become clear in what follows it is convenient to choose a different parametrization. In terms of a variable $\beta^2=(\delta+\zeta-\xi)^2$, $\beta>0$, chosen instead of $\xi$, Eq.~(\ref{Mean1}) reads 
\begin{widetext}
\bea
W_1(\tau)&=&{{RR_1}\over{\sqrt{\pi}}}\int_0^\infty d\eta~\eta^3e^{-\eta^2}\int_0^\infty d\zeta~e^{-R^2(\zeta+\delta)^2+R_1\zeta}
\int^\infty_\zeta d\beta~e^{-R_1\beta}~{{\cos{(2\sqrt{\beta^2+\eta^2}\tau)}}\over{\beta^2+\eta^2}}\nonumber\\
&+&{{RR_1}\over{\sqrt{\pi}}}\int_0^\infty d\eta~\eta^3e^{-\eta^2}\int_0^\infty d\zeta~e^{-R^2(\zeta-\delta)^2-R_1\zeta}
\int_0^\zeta d\beta~e^{R_1\beta}~{{\cos{(2\sqrt{\beta^2+\eta^2}\tau)}}\over{\beta^2+\eta^2}}\nonumber\\
&+&{{RR_1}\over{\sqrt{\pi}}}\int_0^\infty d\eta~\eta^3e^{-\eta^2}\int_0^\infty d\zeta~e^{-R^2(\zeta-\delta)^2-R_1\zeta}
\int^\infty_0 d\beta~e^{-R_1\beta}~{{\cos{(2\sqrt{\beta^2+\eta^2}\tau)}}\over{\beta^2+\eta^2}}\,.
\label{FHM5}
\eea
\end{widetext}
Introducing functions $G_\pm(\zeta)$ of Eq.~(\ref{Trans6}) and performing integrations over $\zeta$ by parts allows eliminating integrations over $\beta$. The procedure is similar to deriving (\ref{Trans7}) - (\ref{Trans9}). Finally
\bea
W_1(\tau)&=&{{RR_1}\over{\sqrt{\pi}}}\int_0^\infty d\eta~\eta^3e^{-\eta^2}\nonumber\\
&\times&\int_0^\infty {{d\zeta}\over{\zeta^2+\eta^2}}~F(\zeta)\cos{(2\sqrt{\zeta^2+\eta^2}\tau)},
\label{FHM6}
\eea
with
\be
F(\zeta)={\tilde G}_+(\zeta)e^{-R_1\zeta}-[G_-(\zeta)e^{R_1\zeta}+G_-(0)e^{-R_1\zeta}].
\label{FHM7}
\ee
Functions ${\tilde G}_\pm(\zeta)$ are defined as complementary to $G_\pm(\zeta)$
\be
{\tilde G}_\pm(\zeta)=\int_0^\zeta du~\exp[-R^2(u\pm\delta)^2\pm R_1u]
\label{FHM8}
\ee
and obey a relation
\be
{\tilde G}_\pm(-\zeta)= - {\tilde G}_\mp(\zeta)\,.
\label{FHM9}
\ee
Eq.~(\ref{FHM6}) is the final expression for the time dependent part of the transition probability.

\subsection{Large $\tau$ asymptotic of $W_1(\tau)$}
\label{sec:Asym2}

Eq.~(\ref{FHM6}) demonstrates the convenience of the parametrization employed in (\ref{FHM5}). Indeed, square root in the argument of $\cos{(2\sqrt{\zeta^2+\eta^2}\tau)}$ can be conveniently eliminated by transforming to polar coordinates as $\zeta=\rho\cos\phi$, $\eta=\rho\sin\phi$. Then (\ref{FHM6}) reads
\bea
&&W_1(\tau)={{RR_1}\over{\sqrt{\pi}}}\int_0^{\pi/2}d\phi~\sin^3\phi\nonumber\\
&\times&\int_0^\infty d\rho~\rho^2e^{-\rho^2\sin^2\phi}F(\rho\cos\phi)~\cos{(2\rho\tau)}\,.
\label{As1}
\eea

The internal integral in (\ref{As1}) is a Fourier transformation, and its asymptotic behavior at large $\tau$ is controlled by the singularities of the integrand. It is analytical for $0\leq\rho<\infty$. When both $R,R_1\neq0$, $F(\zeta)$ decays exponentially with $\zeta$, hence, $F(\rho\cos\phi)$ shows exponential decay for $\phi\neq\pi/2$. The factor $\exp{(-\rho^2\sin^2\phi)}$ decays exponentially for $\phi\neq0$. Therefore, the whole integrand decays exponentially for arbitrary value of $\phi$. To find out the behavior of $F(\zeta)$ near $\zeta=0$, it is convenient to transform (\ref{FHM7}) to an equivalent but a more symmetric form
\bea
F(\zeta)&=&[{\tilde G}_+(\zeta)e^{-R_1\zeta}-{\tilde G}_-(\zeta)e^{R_1\zeta}]\nonumber\\
&-&2G_-(0)\cosh{(R_1\zeta)}\,.
\label{As2}
\eea
It is seen from the comparison of (\ref{FHM9}) and (\ref{As2}) that $F(\zeta)$ is an even function of $\zeta$, $F(-\zeta)=F(\zeta)$, and because it is analytical at $\zeta=0$, all its odd derivatives vanish at $\zeta=0$. This observation allows extending the integration over $\rho$ to the whole real axis, $-\infty<\rho<\infty$, and suggests that the integral vanishes faster than any finite power of $\tau$. Because this statement is valid for arbitrary $\phi$, it is valid also for the integral over $\phi$. Therefore, the decay of $W_1(\tau)$ is of the exponential type, and this is apparently the most general statement that can be made. As distinct from Sec.~\ref{sec:asym1}, I see no way for finding the $\tau\rightarrow\infty$ behavior before performing expansion in inverse $R_1$.

Expanding $F(\zeta)$ in $R_1^{-1}$ provides a more specific outlook onto the asymptotic behavior of $W_1(\tau)$ of (\ref{As1}). Because equations are rather bulky, results will be first provided for the resonance regime, $\delta=0$. The leading term of the expansion of $F(\zeta)$ is $F(\rho\cos\phi)\approx(2/R_1)\exp(-R^2\rho^2\cos^2\phi)$, and by substituting it into (\ref{As1}) one arrives at
\bea
W_1(\tau)&=&{{R}\over{\sqrt{\pi}}}\int_0^{\pi/2}d\phi~\sin^3\phi\int_{-\infty}^\infty d\rho~\rho^2\nonumber\\
&\times&\exp[-\rho^2(\sin^2\phi+R^2\cos^2\phi)]\cos{(2\rho\tau)}.
\label{As3}
\eea
The Gaussian integral over $\rho$ can be performed exactly, but the result is rater cumbersome. It simplifies essentially when terms of the order of 1 are disregarded compared with the term $\tau^2\gg1$. Afterwards, the integration over $\phi$ can be simplified by choosing a proper parametrization. Remarkably, for $R=1$ the integrations over $\rho$ and $\phi$ separate, and $W_1(\tau)$ shows a somewhat special behavior as will be seen from some examples in what follows.

For $R<1$, by choosing 
\be
u=1/[1-(1-R^2)\cos^2\phi]
\label{As3a}
\ee 
as a new variable, one finds
\be
W_1(\tau)\approx-{{R\tau^2}\over{2(1-R^2)^{3/2}}}\int_1^{1/R^2}du~{{1-R^2u}\over{\sqrt{u-1}}}~e^{-u\tau^2}.
\label{As4}
\ee
Because $\exp{(-u\tau^2)}$, in the asymptotic region of large $\tau$, decreases very fast with $u$, the main contribution to the integral comes from the close vicinity of the lower integration limit, $u=1$. However, the integrand in (\ref{As4}) either diverges in this point, for $R\neq1$, or vanishes, for $R=1$. As a result, the asymptotic behavior of the integral is strongly influenced by the competition between $\tau$ and $|R-1|$. For $\tau^2(1-R)\gg1$
\be
W_1(\tau)\approx-{{R\sqrt{\pi}}\over{2\sqrt{1-R^2}}}~\tau e^{-\tau^2}.
\label{As5}
\ee
In the experiments by Laird {\it et al.},\cite{Laird07} the coefficient $R$ was small, $R\approx10^{-2}$. Nevertheless, it is instructive to consider a wider region of $R$ values. Eq.~(\ref{As5}) is valid for $\tau^2(1-R)\gg1$, but in the close vicinity of $R=1$, for $1\ll\tau^2\ll|1-R|^{-1}$, an intermediate asymptotic holds
\be
W_1(\tau)\approx-~{2\over3}~\tau^2e^{-\tau^2}.
\label{As6}
\ee
For $R>1$, the following equation holds instead of (\ref{As4})
\be
W_1(\tau)\approx-{{R\tau^2}\over{2(R^2-1)^{3/2}}}\int_{1/R^2}^1 du~{{R^2u-1}\over{\sqrt{1-u}}}~e^{-u\tau^2},
\label{As7}
\ee
and integration near the lower limit, for $\tau^2(R-1)\gg1$, results in
\be
W_1(\tau)\approx-{{R^2}\over{2(R^2-1)}}~e^{-\tau^2/R^2}.
\label{As8}
\ee

Despite the difference in the specific form of the asymptotic expressions of (\ref{As6}) - (\ref{As8}), they have three major features in common. First, in all cases the decay is of exponential type, close to Gaussian, in agreement with our general conclusion. Second, as distinct from the Rabi oscillations typical of spin-orbit and magnetically mediated EDSR (see Sec.~\ref{sec:SOMM}),\cite{Nowack07,Tarucha08} smooth time dependence is the distinctive feature of hyperfine-mediated EDSR.\cite{Laird07}  Rabi oscillations are washed out by averaging over the fluctuations of the transverse nuclear magnetization. Third, in all cases $W_1(\tau)$ is negative. This suggests that the probability $W(t)$ of (\ref{SOM1}) approaches its limit value, at $t\rightarrow\infty$, from above, in agreement with a weak overshoot found numerically by Laird {\it et al.}\cite{Laird07}

Corrections to (\ref{As3}) from the higher order terms of the expansions of $G_\pm(\zeta)$ and ${\tilde G}_\pm(\zeta)$ in $1/R_1$ contain the same Gaussian exponent $\exp{(-R^2\rho^2\cos^2\phi)}$. Their pre-exponential factors are small in $(R/R_1)^2=(\Xi^2/2\Delta)^2$ and $(R^2\zeta/R_1)^2$ what is actually the same because $R\zeta\alt1$ due to the Gaussian exponent in (\ref{As3}). These corrections can also be estimated in terms of the time $\tau$, after performing both integrations in (\ref{As3}). They are of the order of $(R^2\tau/R_1)^2=[(\Omega/\Delta)(\Xi^2/2\Delta)]^2\tau^2$ and can become important only for very large values of $\tau$ because the estimate includes, in addition to the small parameter $\Xi^2/2\Delta$, Eq.~(\ref{Tr5a}), also a factor $\Omega/\Delta$ that is typically small.

Eq.~(\ref{As3}) was derived for exact resonance, $\delta=0$. For $\delta\neq0$, the leading term of the expansion of $F(\zeta)$ in $1/R_1\ll1$ acquires a factor $\exp{(-\delta^2R^2)}\cos{(2\delta R^2\zeta)}$. After performing the Gaussian integration over $\rho$ in the modified equation (\ref{As3}), the leading term of the expansion in $\tau^2\gg1$ acquires a factor $\cos{(2\delta\tau/\cos{\phi})}$. After changing the variable from $\phi$ to $u$ according to (\ref{As3a}), a factor $\cos{[2\delta\sqrt{u(1-R^2)/(u-1)}~\tau]}$ should be incorporated into the integrand of Eq.~(\ref{As4}). Finally, one arrives at a generalized equation (\ref{As5})
\be
W_1(\tau)\approx-{{R\sqrt{\pi}\exp{[-\delta^2(1-R^2)]}}\over{2\sqrt{1-R^2}}}~\tau e^{-\tau^2}.
\label{As9}
\ee
Remarkably, for $R\ll1$ the suppression of EDSR due to the factor $e^{-\delta^2}$ is much stronger than in (\ref{Asym3}) where the suppression factor $e^{-\delta^2R^2}$. However, for $R=1$ the suppression disappears at all. This is another manifestation of the special EDSR regime near the $R=1$ point where the driving force coincides with the characteristic spectrum width.

One more difference between the off-resonance regime, $\delta\neq0$, and the on-resonance regime, $\delta=0$, is in the structure of the $1/R_1$ expansion series. The expansion coefficient $R^2\tau/R_1$ is the same in both cases, but for $\delta\neq0$ the series includes all powers of $R^2\tau/R_1$, odd and even, while in the resonance regime odd terms vanish.

In all cases the characteristic time of EDSR, when $W(\tau)$ reaches its flat maximum, is $\tau\sim1$ or $t\sim\hbar/\Omega$.

\section{Conclusions}
\label{sec:concl}

As distinct from ferromagnets, where macroscopic magnetization is stabilized by strong exchange interaction, direct interaction of nuclear spins in quantum dots is very weak. Therefore, the description of nuclear polarization as a macroscopic field acting on electron spin is based on a large number of nuclear spins inside the dot, $N\gg1$, and their slow dynamics compared with electron spin dynamics. Meanwhile, the hyperfine coupling of electron spin to nuclear spins is usually much stronger than the direct coupling between nuclear spins, and under the conditions of EDSR the Zeeman term vanishes in the rotating frame. Therefore, applying macroscopic approach to EDSR needs justification. Moreover, only consistent microscopic approach to EDSR can enable calculating corrections to macroscopic theory and provide solid basis for including nuclear dynamics and investigating instabilities in electron-nuclear system.

From the standpoint of {\it the Central Limit Theorem}, one anticipates that EDSR probability can be found by averaging the Rabi formula
\be
W_R(t)=f^2~{{\sin^2{[\sqrt{(\epsilon/2)^2+f^2}~t/\hbar]}}\over{(\epsilon/2)^2+f^2}}
\label{C1}
\ee
over the Gaussian distribution of random nuclear magnetization. However, having in mind that both detuning $\epsilon$ and driving force $f$ depend on nuclear magnetization, performing the averaging of $W_R(t)$ is a challenging problem {\it per se}. Deriving this recipe from microscopic theory and developing techniques for going beyond the macroscopic description is even a much more challenging task.

The approach developed in this paper is based on the following main steps. First, because the expansion of $\sin{({\cal H}t/\hbar)}/{\cal H}$ in power series includes only even powers of $\cal H$, Eq.~(\ref{AV1}), the quantum-mechanical expression for spin-flip probability $W(t)$ can be expressed in terms of polynomials in components of nuclear angular momenta ${\bf I}_j$. Second, retaining only pair correlators (an equivalent of mean field theory), the contributions of the longitudinal and transverse magnetization can be decoupled, and statistical weights of different contributions can be expressed in terms of $\Gamma$-functions of half-integer and integer arguments for longitudinal and transverse magnetizations, respectively. Third, employing integral representations for these $\Gamma$-functions, one arrives at Eq.~(\ref{Mean1}) describing one-dimensional averaging over longitudinal fluctuations and two dimensional averaging over transverse ones. Fourth, higher correlators can be accounted for in a regular way, see Appendix B.

The probability $W(t)$ can be split into its saturation value $W_\infty$ and time dependent part $W_1(t)$, Eq.~(\ref{SOM1}). For spin-orbit and inhomogeneous Zeeman-coupling mediated EDSR, $W_1(t)$  can be reduced to onefold integrals (\ref{Trans7}) - (\ref{Trans9}); its large-$t$ behavior is described by oscillations showing a $t^{-1/2}$ decay in agreement with Ref.~\onlinecite{Koppens07}. For hyperfine mediated EDSR, time dependence is close to monotonic, with a weak overshoot as found in Ref.~\onlinecite{Laird07}, and is described by a twofold integral (\ref{As1}). Large-$t$ asymptotic is Gaussian in the framework of mean field theory, but it develops a power-law tail when the first four-spin correlator correction is accounted for, see Appendix B. The contribution of such a tail should become essential at moderate times $t\propto \ln{N}$, where $N$ is the number of nuclei in the dot.

Regarding the effect of nuclear spin-spin coupling on EDSR, it is of critical importance that this coupling shows no resonant enhancement under the EDSR conditions, see Appendix A. As a result, it has no effect on EDSR as long as the criterion (\ref{Tr5a}) is fulfilled. Its explicit form is
\be
{{\Xi^2}\over{2\Delta}}=\sqrt{{{(An_0)^2I(I+1)}\over{8N\hbar\omega_0\hbar\omega_s}}}\,,\,\,N=\pi r_0^2dn_0\,.
\label{C2}
\ee
For GaAs quantum dots with $An_0\approx0.1$ meV, $\hbar\omega_0\approx0.1$ meV, $d\approx5$ nm, and $B\approx100$ G, this ratio is small, ${{\Xi^2}/{2\Delta}}\sim10^{-2}\ll1$.

It is the main restriction of the model employed in this paper that it includes only ``passive" processes in the nuclear subsystem like spin relaxation, and does not include mechanisms that can result in pumping nuclear polarization by electron dynamics. Including such ``active" processes and related instabilities in the coupled electron-nuclear system into the formalism developed above is the major challenge.

\acknowledgments

I am grateful to E. A. Laird, C. Barthel, and C. M. Marcus, whose discovery of hyperfine-mediated EDSR inspired this research, for fruitful collaboration, to Al. L. Efros and J. J. Krich for clarifying discussions, and to B. I. Korenblum for his insights and advice regarding the asymptotic behavior of integrals. The support from the Center for Nanoscale Systems of Harvard University and from the Loughborough University through the Rutherford Professorship is gratefully acknowledged.

\appendix
\section{Dynamic nuclear spin-spin coupling}

Eq.~(\ref{spsp}) was  derived for a static Hamiltonian $H_Z+H_{hf}^\perp$. For exploring the role of driving force on spin-spin coupling, one can use a model Hamiltonian
\bea
H=-{1\over2}\hbar\omega_s\sigma_z
&+&{1\over2}\sqrt{{{\hbar\omega_s}\over{2}}}\sum_ja_j(I_j^+\sigma_-+I_j^-\sigma_+)\nonumber\\
&+&{1\over2}~f(\sigma_+e^{i\omega t}+\sigma_-e^{-i\omega t})\,.
\label{A1}
\eea
Two first terms are identical to $H_Z+H_{hf}^\perp$\,, with the second term rewritten in terms of the coefficients $a_j$ of Eq.~(\ref{aDefin}). Last term describes the driving force in the rotating wave approximation, with $f$ defined as $f=f_\pm$ of Eq.~(\ref{RF3}). The corresponding rotating frame Hamiltonian is
\bea
{\hat H}&=&{\hat H}_0+h(t)\,,\,\,{\hat H}_0={1\over2}\Delta_s\sigma_z+f\sigma_x\nonumber\\
h(t)&=&\sqrt{{{\hbar\omega_s}\over{8}}}\sum_ja_j(I_j^+\sigma_-e^{i\omega t}+I_j^-\sigma_+e^{-i\omega t}).
\label{A2}
\eea
For finding a dynamic analog of (\ref{spsp}), effect of $h(t)$ on slow nuclear dynamics should be calculated in the second-order perturbation theory. To this end, by rotation in the spin space
\be
\sigma_x={{\sigma_1-\gamma\sigma_3}\over{\sqrt{1+\gamma^2}}}\,,\,\,\sigma_y=\sigma_2\,,\,\,
\sigma_z={{\sigma_3+\gamma\sigma_1}\over{\sqrt{1+\gamma^2}}}\,,
\label{A3}
\ee
with $\gamma=\Delta_s/2f$, the Hamiltonian ${\hat H}_0$ simplifies as
\be
{\hat H}_0=f^\prime\sigma_1\,,\,\,f^\prime=f\sqrt{1+\gamma^2}\,,
\label{eqA4}
\ee
and $h(t)$ can be transformed into the interaction representation ${\tilde h}(t)=\exp{(i{\hat H}_0t/\hbar)}h(t)\exp{(-i{\hat H}_0t/\hbar)}$ by using
\bea
\sigma_1(t)=\sigma_1\,,\,\,\sigma_2(t)&=&\sigma_2-\sigma_3\sin{(2f^\prime t)}\,,\nonumber\\
\sigma_3(t)&=&\sigma_3+\sigma_2\sin{(2f^\prime t)}\,;
\label{eqA5}
\eea
here and below, $f^\prime$ in frequency units. Then, calculating the second order correction through a $t$-ordered product
\be
{{(-i)^2}\over{2!}}\int_0^t\int_0^t T\{{\tilde h}(t_1){\tilde h}(t_2)\}dt_1dt_2\,,
\label{A6}
\ee
eliminating all fast-oscillating terms proportional to $e^{\pm i\omega t}$, and exponentiating the result, one finds a hyperfine correction to the phase of the electron-spin wave function. Representing the phase factor as $\exp[-iH_{ss}(t)t/\hbar]$ allows finding effective spin-spin coupling $H_{ss}(t)$.

Exact expression for $H_{ss}(t)$ is rather bulky, but in the large $\omega$ limit it essentially simplifies
\bea
&&H_{ss}(t)=-{1\over2}{{\omega_s/\omega}\over{\sqrt{1+\gamma^2}}}
\sum_{mn}a_ma_nI_m^+I_n^-\nonumber\\
&\times&\bigg[{{\gamma}\over{2}}\bigg(3-{{\sin{(4f^\prime t)}}\over{4f^\prime t}}\bigg)\sigma_1+{{\sin^2{(f^\prime t)}}\over{f^\prime t}}\sigma_2+\sigma_3\bigg].
\label{A7}
\eea
Near the spin resonance $\omega\approx\omega_s$, and the factor $\omega_s/\omega\approx1$ can be omitted. $H_{ss}(t)$ oscillates with the Rabi frequency $f^\prime$ and saturates for $f^\prime t\rightarrow\infty$. Expressing $\{\sigma_1,\sigma_3,\sigma_3\}$ through $\{\sigma_x,\sigma_y,\sigma_z\}$ by using (\ref{A3}), one can check that at short times, $f^\prime t\ll1$, $H_{ss}(t)$ coincides with $H_{ss}$ of Eq.~(\ref{spsp}). In the opposite limit, for $\gamma\gg1$ and $f^\prime t\gg1$, $H_{ss}(t)\approx3H_{ss}/2$.

Therefore, $H_{ss}(t)$ shows no resonant enhancement at $\omega\rightarrow\omega_s$ and has a magnitude comparable to $H_{ss}$ of Eq.~(\ref{spsp}) and the same sign. In physical terms, this result suggests that the hyperfine dynamics near the resonance, $\Delta_s\approx0$, is strong enough to smear the rotating frame singularity and reduce its effect to moderate changes in the nuclear spin relaxation rate that acquires oscillatory behavior. Regarding the significance of this conclusion, see discussion at the end of Sec.~\ref{sec:spsp}.

\section{Effect of higher correlators}

Eq.~(\ref{Mean1}) was derived including only pair correlators of nuclear spins $\langle I_m^\alpha I_n^\beta\rangle=\langle I_\alpha I_\beta\rangle\delta_{mn}$, Sec.~\ref{sec:PairCor}. In this Appendix, effect of higher correlators on EDSR intensity will be estimated as applied to the operators $f_\pm^{hf}$ of Eq.~(\ref{RFhf}). To simplify bulky calculations, terms $g_z$ and $g_+g_-$ in the Hamiltonian of Eq.~(\ref{RF3}) will be omitted.

In the pair-correlator approximation, a correlator $F_M=\langle(f_+f_-)^M\rangle$ including $M$ pairs of operators $f_\pm$, was evaluated as
\be
{\cal F}_M= M!\langle I_+I_-\rangle^M(\sum_mb_m^2)^M\,.
\label{B1}
\ee
This estimate came from counting the number of possible $I_m^+I_n^-$ pairings without any constraints imposed on $m$ and $n$. However, such a count overestimates the contribution of pair-correlators that should be found from
\be
F^{(0)}_M=M!<I_-I_+>^M\sum_{m_1\neq m_2\neq...\neq m_M}b_{m_1}^2b_{m_2}^2...b_{m_M}^2\,.
\label{B2}
\ee
Indeed, each sharing of indices like $m_1=m_2$ results in higher correlators like $\langle (I_+I_-)^2\rangle$. Therefore, one needs to evaluate the effect of this constraint on $W(t)$. For this evaluation, it is enough to keep in (\ref{B2}) only the principle sequence of corrections that includes terms with a number of pairs of shared indices, like $m_1=m_2$, but does not include triple shared indices, like $m_1=m_2=m_3$, and terms with a larger number of shared indices. Their contribution is of higher order in $1/N$, where $N=\pi r_0^2dn_0$ is the effective number of nuclei in the dot.

Each term of the correlator ${\cal F}_M^{(0)}$ involving $s$ pairs of shared indices $m_j$, makes a contribution
\be
(-)^s{\cal F}_M~{{M!}\over{s!2^s(M-2s)!}}(\sum_mb_m^4)^s/(\sum_mb_m^2)^{2s}\,.
\label{B3}
\ee
Here $0\leq s\leq s_{\rm max}$, with $s_{\rm max}=M/2$ when $M$ is even and $s_{\rm max}=(M-1)/2$ when $M$ is odd. The factor with factorials equals the multiplicity of independent selections of $s$ pairs of shared indices $m_j$. The origin of the sign factor $(-)^s$ becomes clear when one eliminates, in succession, first a single pair of shared indices, then the second, etc. Last factor can be calculated using the definition (\ref{RFhf}) of $b_m$ with $\Psi_0({\bf r},y)$ of Sec.~\ref{sec:Long}
\be
\sum_mb_m^4/(\sum_mb_m^2)^2=35/24N\,.
\label{B4}
\ee
Therefore, in the principle sequence approximation, $F^{(0)}_M$ equals to
\be
F^{(0)}_M\approx{\cal F}_M\sum_{s=0}^{s_{\rm max}}(-)^s{{M!}\over{s!(M-2s)!}}\left({{35}\over{48N}}\right)^s.
\label{B5}
\ee
Estimating a few first terms of the series shows that the expansion parameter is $M^2/N$. It corresponds to a typical fluctuation $M\sim\sqrt{N}$ in the nuclear reservoir. With $M\gg1$ and $s\ll M$, the ratio $M!/(M-2s)!\approx M^{2s}$, and (\ref{B5}) is an exponential series. It converges fast, and one can extend summation to infinity. Then
\be
F_M^{(0)}\approx{\cal F}_M\exp{(-35M^2/48N)}\,.
\label{B6}
\ee
Therefore, constraint (\ref{B2}) results in an exponential factor of Eq.~(\ref{B6}).

For calculating the effect of the constraint (\ref{B2}) on $W(\tau)$, Eqs.~(\ref{B1}) and (\ref{B6}) should be plugged into (\ref{AV2}) (with $g_z, g_\pm=0$), and $M!$ should be eliminated by applying averaging over the Gaussian distribution of Eq.~(\ref{Tr8}). Then one arrives at
\bea
W^{(0)}(t)&\approx&\int d^2{\bf v}~ \rho_\Omega(v)\sum_{k,k'=0}^\infty{{(-)^{k+k'}(t/\hbar)^{2(k+k'+1)}}\over{(2k+1)!(2k'+1)!}}\nonumber\\
&\times&\sum_{\nu=0}^{k+k'}{{(k+k')!}\over{\nu!(k+k'-\nu)!}}\biggl({{\Delta_s}\over{2}}\biggr)^{2\nu}\nonumber\\
&\times&\bigg[v^{2M}\exp\bigg(-{{35M^2}\over{48N}}\bigg)\bigg]_{M=k+k'-\nu+1}\,,
\label{B7}
\eea
where the superscript zero indicates that $W^{(0)}$ includes only pair correlators. With the last exponential factor omitted, the sum over $\nu$ is a binomial $v^2[(\Delta_s/2)^2+v^2]^{k+k^\prime}$, and (\ref{B7}) reduces to a simplified version of (\ref{Mean1}) with only a single integration.

In resonance, $\Delta_s=0$, the only characteristic time scale is $\hbar/\Omega$. For $\tau=t\Omega/\hbar\sim1$, sums over $k$ converge fast, at $k, k^\prime \sim1$. Therefore, $M\sim1$ and the effect of the exponential factor $\exp(-{{35M^2}/{48N}})$ is small, only about $1/N$. For large detuning, $\Delta_s\gg\Omega$, a short time scale $\hbar/\Delta_s$ develops. However, it was shown in Sec.~\ref{sec:Asym2} that EDSR frequency fluctuations $\Delta/\hbar$ of a scale $\Delta\agt\Delta_s/2$ [that were omitted when deriving (\ref{B7})] reestablish the $\hbar/\Omega$ scale, hence, the $1/N$ estimate should remain valid for the hyperfine-mediated EDSR.

The difference between $F_M^{(0)}$ and ${\cal F}_M$ reflected the reduction in the contribution of two-spin correlators, for given $M$, because of the constraint (\ref{B2}). Next step is estimating the direct contribution of higher correlators, and this will be done for $\Delta_s=0$. The contribution of a single four-spin correlator $\langle(I_+I_-)^2\rangle$ to $W(t)$ equals
\bea
W^{(0)}(t)&\approx&\sum_{k,k'=0}^\infty{{(-)^{k+k'}(t/\hbar)^{2(k+k'+1)}}\over{(2k+1)!(2k'+1)!}}\nonumber\\
&\times&{{7}\over{4N}}\left[{\cal F}_M{{M(M-1)}\over{4}}\right]_{M=k+k^\prime+1}\,.
\label{B8}
\eea
The coefficient is a product of the ratio $\langle(I_+I_-)^2\rangle/\langle I_+I_-\rangle^2=6/5$ evaluated for classical spins and the ratio of $b$-sums of Eq.~(\ref{B4}). The contribution of pair correlators was estimated in the principle sequence approximation, similarly to (\ref{B3}). The multiplier $M(M-1)/4$ reflects change in the combinatorial factor due to adding a single four-spin correlator and reducing the number of two-spin correlators by two (total number of spin operators kept fixed). Exponential factor about $\exp{(-36M^2/48N)}$ is omitted.

The factorial $M!$ in ${\cal F}_M$ can be eliminated by introducing a Gaussian variable $v$ and performing Gaussian integration according to (\ref{Tr8}). Afterwards, the factor $M(M-1)$ can be eliminated in the integrand by substituting it by derivatives like $Mv^M=vd(v^M)/dv$. This transformation allows performing summations over $k$ and $k^\prime$, and one arrives at
\bea
&&W^{(1)}(t)\approx{{7}\over{8N\Omega^2}}\int_0^\infty dv~v^3e^{-v^2/\Omega^2}\nonumber\\
&\times&\left[\left(v^2{{d}\over{d(v^2)}}\right)^2+v^2{{d}\over{d(v^2)}}\right]{{\sin^2(vt/\hbar)}\over{v^2}}\,.
\label{B9}
\eea
The integral can be simplified by performing by parts. Finally, in terms of $\tau=t\Omega/\hbar$,
\be
W^{(1)}(\tau)={{7}\over{8N}}\int_0^\infty du~u(2-4u^2+u^4)e^{-u^2}\sin^2{u\tau}\,.
\label{B10}
\ee
Its asymptotic behavior for $\tau\gg1$
\be
W^{(1)}(\tau)\approx 7/(16N\tau^2)\,;
\label{B11}
\ee
the power-law decay originates due to finite slope of the integrand at $u=0$. Therefore, at the time span of $\tau\sim1$, first correction to mean-field behavior of $W(t)$ from the four-spin correlator is of the order of $1/N$, i.e., of the same order as the correction (\ref{B7}) from the reduction of the number of pair correlators. A new feature is the time dependence of $W^{(1)}(\tau)$. While the Gaussian asymptotic of Eqs.~(\ref{As5})-(\ref{As8}) is a distinctive property of the mean-field theory of hyperfine-mediated EDSR, (\ref{B11}) indicates a much slower decay of the corrections to this theory. The power-law asymptotic deserves a more detailed study, however, a cancellation of power-law terms and return to the exponential asymptotic behavior seems improbable. And power-law tails suggest deviations from the mean-field behavior at relatively modest times of about $\tau\sim\ln{N}$.

It deserves mentioning that (\ref{B7}) has implications beyond the simplified model of the hyperfine-mediated EDSR for which it was derived. Spin-orbit and magnetically mediated EDSR's show oscillatory behavior, Sec.~\ref{sec:asym1}, and because the convergence of the series describing oscillations slows down rather fast with the number of oscillations, this immediately translates in the increase in the corrections that are proportional to $M^2\approx(k+k^\prime)^2$ according to (\ref{B7}). Therefore, the magnitude of the corrections to mean field theory is expected to increase essentially with the number of oscillations.


\begin{thebibliography}{99}
\bibitem{SpinRel} For the same reason, spin relaxation in quantum dots is suppressed in the $B\rightarrow0$ limit for all time-inversion conserving spin relaxation mechanisms.\cite{KhaNaz00}
\bibitem{DrRa} The ratio $\alpha_D/\alpha_R$ is material and structure dependent,\cite{PP95,Mil03,Gan04,Ensslin07} but for our problem only their combination $\alpha$ is essential.
\end{thebibliography}
\end{document}